\pdfoutput=1
\documentclass[aip,amsmath,amssymb,reprint,onecolumn]{revtex4-1}

\usepackage{graphicx,hyperref}

\newcommand{\uphi}{\hat{\boldsymbol{\phi}}}
\newcommand{\uR}{\hat{\mathbf{R}}}
\newcommand{\uZ}{\hat{\mathbf{Z}}}
\newcommand{\relphantom}[1]{\phantom{\mathrel{#1}}}

\begin{document}


\title{Multi-region relaxed magnetohydrodynamics with flow} 



\author{G.R. Dennis}
\email[]{graham.dennis@anu.edu.au}
\affiliation{Research School of Physics and Engineering, Australian National University, ACT 0200, Australia}

\author{S.R. Hudson}
\affiliation{Princeton Plasma Physics Laboratory, PO Box 451, Princeton, NJ 08543, USA}

\author{R.L. Dewar}
\author{M.J. Hole}
\affiliation{Research School of Physics and Engineering, Australian National University, ACT 0200, Australia}


\date{\today}

\begin{abstract}

We present an extension of the multi-region relaxed magnetohydrodynamics (MRxMHD) equilibrium model that includes plasma flow.  This new model is a generalization of Woltjer's model of relaxed magnetohydrodynamics equilibria with flow.  We prove that as the number of plasma regions becomes infinite our extension of MRxMHD reduces to ideal MHD with flow.  We also prove that some solutions to MRxMHD with flow are not time-independent in the laboratory frame, and instead have 3D structure which rotates in the toroidal direction with fixed angular velocity.  This capability gives MRxMHD potential application to describing rotating 3D MHD structures such as `snakes' and long-lived modes.

\end{abstract}


\maketitle 

\section{Introduction}

The construction of magnetohydrodynamic (MHD) equilibria in three-dimensional (3D) configurations is of fundamental importance for understanding toroidal magnetically confined plasmas.  The theory and numerical construction of 3D equilibria is complicated by the fact that toroidal magnetic field without a continuous symmetry are generally a fractal mix of islands, chaotic field lines, and magnetic flux surfaces.  \citet{Hole:2007} have proposed a mathematically rigorous model for 3D MHD equilibria without flow that embraces this structure by abandoning the assumption of continuously nested flux surfaces usually made when applying ideal MHD.  Instead a finite number of flux surfaces are assumed to exist in a partially relaxed plasma system.  This model, termed a multi-region relaxed MHD (MRxMHD) model, is based on a generalization of the Taylor relaxation model \citep{Taylor:1974,Taylor:1986} in which the total energy (field plus plasma) is minimized subject to a finite number of magnetic flux, helicity and thermodynamic constraints.  

The MRxMHD model has seen some recent success in describing the 3D quasi-single-helicity states in RFX-mod \citep{Dennis:2013b}, however it must be extended to include plasma flow as rotation and velocity shear play important roles in high-performance devices \citep{Rice:2007}.  Our extension to include flow is guided by the work of \citet{Woltjer:1958}, and \citet{Finn:1983} who studied models for relaxed flowing plasmas by constraining flow helicity $C = \int \mathbf{B} \cdot \mathbf{u}\, d^3\tau$ and angular momentum in addition to the flux and magnetic helicity constraints considered by Taylor \citep{Taylor:1986}.  The models studied by Woltjer, Finn \& Antonsen are the single plasma-region limit of the MRxMHD model with flow presented in this paper.

In the opposite limit, as the number of plasma interfaces becomes large and the plasma contains continuously nested flux surfaces, it is desirable for MRxMHD with flow to reduce to ideal MHD with flow.  We prove this limit to be true in Section~\ref{sec:ContinuousLimit}, demonstrating that MRxMHD with flow essentially `interpolates' between Taylor-Woltjer relaxation theory on the one hand and ideal MHD with flow on the other.  

One of the intriguing features of our model is that it allows the description of plasmas with rotating 3D MHD structures using a minimum-energy approach.  While these plasma states are not time-independent in the laboratory frame, they are in a rotating reference frame, and are in force-balance in that frame.  This property gives MRxMHD potential application to describing rotating 3D structures such as `snakes' \citep{Weller:1987,Pecquet:1997,Menard:2005,Delgado-Aparicio:2013} and long-lived modes \citep{Chapman:2010,Hua:2010}.

This paper is structured as follows: in Section~\ref{sec:MRXMHDWithFlow} we give a summary of the MRxMHD model and its solution for a finite number of plasma regions before presenting our extension to include plasma flow and discussing the effect of flow on relaxed plasma equilibria.  In Section~\ref{sec:ContinuousLimit} we prove that this extension of MRxMHD reduces to ideal MHD with flow in the limit of continuously nested flux surfaces.  This is followed by an example application of the MRxMHD with flow model to an RFP-like plasma in Section~\ref{sec:Example}.  The paper is concluded in Section~\ref{sec:Conclusion}.

\section{The Multi-region relaxed MHD model}
\label{sec:MRXMHDWithFlow}

\subsection{The zero-flow limit}
\label{sec:ZeroFlowMRXMHD}
\begin{figure}
  \includegraphics[width=8cm]{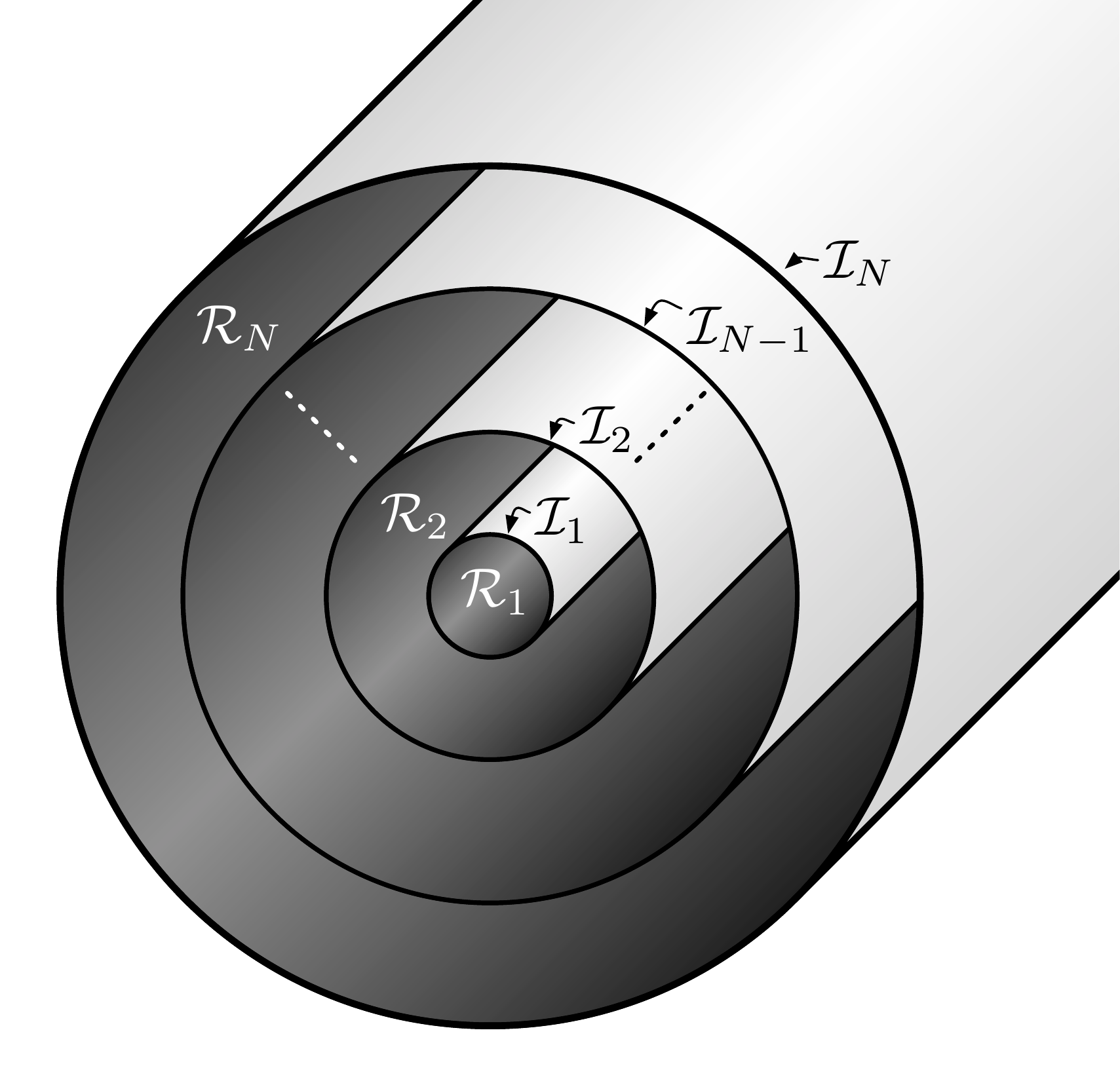}
  \caption{\label{fig:NestedSurfaces}Schematic of magnetic geometry showing ideal MHD barriers $\mathcal{I}_i$, and the relaxed plasma regions $\mathcal{R}_i$.}
\end{figure}

The model we present in this paper is an extension of the MRxMHD model introduced previously \citep{Hole:2006,Hole:2007,Hudson:2007,Dewar:2008}.  Briefly, the MRxMHD model consists of $N$ nested plasma regions $\mathcal{R}_i$ separated by ideal MHD barriers $\mathcal{I}_i$ (see Fig.~\ref{fig:NestedSurfaces}).  Each plasma region is assumed to have undergone Taylor relaxation \citep{Taylor:1986} to a minimum energy state subject to conserved fluxes and magnetic helicity.  The MRxMHD model minimizes the plasma energy
\begin{align}
  E &= \sum_i E_i = \sum_i \int_{\mathcal{R}_i} \left(\frac{1}{2} \mathbf{B}^2 + \frac{1}{\gamma-1}\sigma_i \rho^\gamma \right)\, d^3\tau, & \text{(Plasma Energy)} \label{eq:NoFlowPlasmaEnergy}\\
  \intertext{subject to constraints on the quantities}
  M_i &= \int_{\mathcal{R}_i} \rho\, d^3\tau, & \text{(Plasma Mass)} \label{eq:NoFlowPlasmaMass}\\
  K_i &= \int_{\mathcal{R}_i} \mathbf{A}\cdot\mathbf{B}\, d^3\tau - \Delta\psi_{p,i} \oint_{\mathcal{C}_{p,i}^{<}} \mathbf{A} \cdot d\mathbf{l} - \Delta \psi_{t,i} \oint_{\mathcal{C}_{t,i}^{>}} \mathbf{A} \cdot d\mathbf{l}, & \text{(Magnetic Helicity)} \label{eq:NoFlowMagneticHelicity}
\end{align}
where $\sigma_i = p/\rho^\gamma$, $p$ is the plasma pressure, $\rho$ is the plasma mass density, $\mathbf{A}$ is the magnetic vector potential, and the loop integrals in Eq.~\eqref{eq:NoFlowMagneticHelicity} are required for gauge invariance.  Additionally, each plasma region $\mathcal{R}_i$ is bounded by magnetic flux surfaces and is constrained to have enclosed toroidal flux $\Delta\psi_{t,i}$ and poloidal flux $\Delta\psi_{p,i}$.  The $\mathcal{C}_{p,i}^<$ and $\mathcal{C}_{t,i}^>$ are circuits about the inner ($<$) and outer ($>$) boundaries of $\mathcal{R}_i$ in the poloidal and toroidal directions, respectively.

Minimum energy states of the MRxMHD model are stationary points of the energy functional
\begin{align}
  W &= \sum_i E_i - \sum_{i} \nu_i \left(M_i - M_i^0\right) - \frac{1}{2} \sum_{i} \mu_i \left(K_i - K_i^0\right) , \label{eq:NoFlowEnergyFunctional}
\end{align}
where $\nu_i$ and $\mu_i$ are Lagrange multipliers respectively enforcing the plasma mass and magnetic helicity constraints, and the $M_i^0$ and $K_i^0$ are respectively the constrained values of the plasma mass and magnetic helicity.  

Setting the first variation of Eq.~\eqref{eq:NoFlowEnergyFunctional} to zero gives \citep{Hole:2006}
\begin{align}
  \mathcal{R}_i&: & \nabla \times \mathbf{B} &= \mu_i \mathbf{B}, \label{eq:NoFlowVolumeCondition1}\\
  \mathcal{R}_i&: & p_i &= \text{const},  \label{eq:NoFlowVolumeCondition2}\\
  \mathcal{I}_i&: & \left[\left[p_i + \frac{1}{2} \mathbf{B}^2 \right]\right] &= 0, \label{eq:NoFlowInterfaceCondition}
\end{align}
where Eqs.~\eqref{eq:NoFlowVolumeCondition1}--\eqref{eq:NoFlowVolumeCondition2} apply in each plasma region $\mathcal{R}_i$, Eq.~\eqref{eq:NoFlowInterfaceCondition} applies on each ideal interface $\mathcal{I}_i$, and $[[x]] = x_{i+1} - x_{i}$ denotes the change in quantity $x$ across the interface $\mathcal{I}_i$.

\subsection{Including the effects of plasma flow}
\label{sec:MRxMHDWithFlow}

We present here an extension to MRxMHD to include the effects of plasma flow.  In this model each plasma region is assumed to have undergone a generalized type of Taylor relaxation which minimizes the plasma energy\footnote{As originally discussed by \citet{Kruskal:1958}, minimizing the plasma energy requires invoking an unspecified mechanism that removes the excess energy from the plasma.  An alternative approach, following \citet{Finn:1983} would be to instead maximize the plasma entropy while conserving the plasma energy and other constraints in each volume.}
\begin{align}
  E &= \sum_i E_i = \sum_i\int_{\mathcal{R}_i} \left(\frac{1}{2}\rho \mathbf{u}^2 + \frac{1}{2} \mathbf{B}^2 + \frac{1}{\gamma-1}\sigma_i \rho^\gamma \right)\, d^3\tau & \text{(Plasma Energy)} \label{eq:FiniteVolumePlasmaEnergy}\\
  \intertext{subject to constraints on the plasma mass (Eq.~\eqref{eq:NoFlowPlasmaMass}), magnetic helicity (Eq.~\eqref{eq:NoFlowMagneticHelicity}), and the additional quantities}
  C_i &= \int_{\mathcal{R}_i} \mathbf{B}\cdot\mathbf{u}\, d^3\tau, & \text{(Flow Helicity)} \label{eq:FiniteVolumeFlowHelicity}\\
  L_{i} &= \uZ \cdot \int_{\mathcal{R}_i} \rho \mathbf{r} \times \mathbf{u}\, d^3\tau = \int_{\mathcal{R}_i} \rho R \mathbf{u} \cdot \uphi\, d^3\tau, & \text{(Toroidal Angular Momentum)} \label{eq:FiniteVolumeAngularMomentum}
\end{align}
where $\sigma_i = p/\rho^\gamma$, $\rho$ is plasma mass density, $\mathbf{u}$ is the mean plasma velocity, and $\mathbf{A}$ is the magnetic vector potential.  The plasma quantities constrained by MRxMHD with flow are all conserved by ideal MHD, and are assumed to be robust in the presence of small amounts of resistivity and viscosity.  \citet{Qin:2012} have recently proven this to be true for the magnetic helicity $K$, and our choice to constrain the flow helicity $C$ is motivated by the work of \citet{Woltjer:1958} and \citet{Finn:1983}.  Whether or not the toroidal angular momentum constraint should be enforced will depend on the symmetry of the problem.  This issue is discussed in detail in Section~\ref{sec:AngularMomentumConstraint}.

Minimum energy states of the MRxMHD model with flow given by Eqs.~\eqref{eq:FiniteVolumePlasmaEnergy}--\eqref{eq:FiniteVolumeAngularMomentum} are stationary points of the energy functional
\begin{align}
  W &= \sum_i E_i - \sum_i \nu_i(M_i - M_i^0) - \frac{1}{2}\sum_i \mu_i(K_i - K_i^0) - \sum_i \lambda_i (C_i - C_i^0) - \sum_i \Omega_i (L_{i} - L_{i}^0), \label{eq:FiniteVolumeEnergyFunctional}
\end{align}
where the $\nu_i$, $\mu_i$, $\lambda_i$ and $\Omega_i$ are Lagrange multipliers enforcing constraints on the quantities in Eqs.~\eqref{eq:FiniteVolumePlasmaEnergy}--\eqref{eq:FiniteVolumeAngularMomentum}, and the $X_i^0$ are the constrained values of the quantities $X_i$.

Setting the first variation of Eq.~\eqref{eq:FiniteVolumeEnergyFunctional} to zero gives the plasma region conditions
\begin{align}
  \mathcal{R}_i: &&\nabla \times \mathbf{B} &= \mu_i \mathbf{B} + \lambda_i \nabla \times \mathbf{u},  \label{eq:FiniteVolumeField}\\
  \mathcal{R}_i: &&\rho \mathbf{u} &= \lambda_i \mathbf{B} + \rho \Omega_i R \uphi, \label{eq:FiniteVolumeFlow}\\
  \mathcal{R}_i: &&\nu_i &= \frac{1}{2}\mathbf{u}^2 + \frac{\gamma}{\gamma - 1} \sigma_i \rho^{\gamma - 1} - \Omega_i R \mathbf{u}\cdot\uphi, \label{eq:FiniteVolumeBernoulli} \\
\intertext{and the interface condition}
  \mathcal{I}_i: && \left[\left[\frac{1}{2} \mathbf{B}^2 + p\right]\right] &= 0. \label{eq:InterfaceForceBalance}
\end{align}
The plasma region conditions Eqs.~\eqref{eq:FiniteVolumeField}--\eqref{eq:FiniteVolumeBernoulli} are identical to those derived previously \citep{Woltjer:1958,Finn:1983} in the case of a single relaxed region.  The interface condition, Eq.~\eqref{eq:InterfaceForceBalance}, is the same as Eq.~\eqref{eq:NoFlowInterfaceCondition} for MRxMHD without flow.  A derivation of Eqs.~\eqref{eq:FiniteVolumeField}--\eqref{eq:InterfaceForceBalance} is given in Appendix~\ref{sec:MRxMHDDerivation}.

The MRxMHD with flow model described here reduces to the no-flow limit presented in the previous section if the flow helicity and angular momentum constraints are relaxed.  In this limit the Lagrange multipliers $\lambda_i$ and $\Omega_i$ are zero and Eqs.~\eqref{eq:FiniteVolumeField}--\eqref{eq:InterfaceForceBalance} reduce to Eqs.~\eqref{eq:NoFlowVolumeCondition1}--\eqref{eq:NoFlowInterfaceCondition}.

We validate our model in Section \ref{sec:ContinuousLimit} by proving that it approaches ideal MHD in the limit as the number of plasma volumes $N$ becomes large.  We have previously proven in the absence of flow that MRxMHD approaches ideal MHD without flow \citep{Dennis:2013a}.

\subsection{The toroidal angular momentum constraint}
\label{sec:AngularMomentumConstraint}

If the plasma boundary is axisymmetric, then the total toroidal angular momentum within the plasma $L = \sum_i L_i$ will be conserved. If the boundary is not axisymmetric, torque can be exerted on the plasma even if it is ideal\citep{Taylor:2003}.  This also holds for each plasma region individually: if the interfaces of a given plasma region are initially axisymmetric and remain axisymmetric during the relaxation process, then its toroidal angular momentum $L_i$ will be conserved.  Thus enforcing the conservation of toroidal angular momentum $L_i$ within each plasma volume is equivalent to assuming that all plasma interfaces remain axisymmetric during the plasma relaxation process.  This is a very strong assumption that will be appropriate for some plasmas, but inappropriate for others, for example rotating fully 3D MHD structures such as `snakes' and long-lived modes.

A more appropriate model for rotating 3D MHD structures is obtained if instead of enforcing the conservation of toroidal angular momentum $L_i$ in each region individually, only the total toroidal angular momentum $L = \sum_i L_i$ is conserved.  This only requires the assumption that the outer plasma boundary be axisymmetric, which is reasonable for snakes and long-lived modes.  The equations equivalent to Eqs.~\eqref{eq:FiniteVolumeEnergyFunctional}--\eqref{eq:InterfaceForceBalance} for this situation can simply be obtained by making the replacement $\Omega_i \rightarrow \Omega\; \forall i$ as this replacement makes Eq.~\eqref{eq:FiniteVolumeEnergyFunctional} the appropriate energy functional for the conservation of the total toroidal angular momentum $L$.

Finally, the toroidal angular momentum constraint must be completely relaxed if the outer plasma boundary is not axisymmetric (e.g.\ stellarators).  In this case the toroidal angular momentum is not conserved because normal forces exerted by the non-axisymmetric boundary on the plasma can exert a non-zero torque.  In the next section we show that a contradiction would arise if the toroidal angular momentum constraint was assumed to hold for plasmas with non-axisymmetric boundaries.  Relaxing the toroidal angular momentum constraint entirely is achieved in the energy functional Eq.~\eqref{eq:FiniteVolumeEnergyFunctional}, and Eqs.~\eqref{eq:FiniteVolumeField}--\eqref{eq:InterfaceForceBalance} by making the replacement $\Omega_i \rightarrow 0$.

\subsection{The effects of flow on MRxMHD}

\label{sec:MRxMHDFlowEffects}

An important consequence of including flow in MRxMHD is that the plasma minimum energy states may no longer be time-independent in the laboratory reference frame.  This can be seen by using Eqs.~\eqref{eq:FiniteVolumeField}--\eqref{eq:FiniteVolumeFlow} to show that the minimum energy states of MRxMHD with flow obey the equation
\begin{align}
  \rho \left(\mathbf{u} \cdot \nabla \right) \mathbf{u} &= -\nabla p + \mathbf{J} \times \mathbf{B} - \rho \Omega_i R \uphi \times (\nabla \times \mathbf{u} ) + \rho \Omega_i \nabla (R \mathbf{u}\cdot\uphi). \label{eq:MRxMHDForceEquation}
\end{align}
Comparing this to the evolution equation for ideal MHD with flow
\begin{align}
  \rho \frac{\partial}{\partial t} \mathbf{u} + \rho (\mathbf{u} \cdot \nabla) \mathbf{u} &= - \nabla p + \mathbf{J} \times \mathbf{B}, \label{eq:IdealMHDEvolution}
\end{align}
demonstrates that the minimum energy state described by Eq.~\eqref{eq:MRxMHDForceEquation} will not in general be time-independent in the laboratory frame unless the last two terms of Eq.~\eqref{eq:MRxMHDForceEquation} are zero.  Each plasma region is, however, time-independent in a reference frame rotating about the $Z$ axis with angular frequency $\Omega_i$.  This is seen by making the replacement $\mathbf{u} = \mathbf{u}' + \Omega_i R \uphi$ where $\mathbf{u}'$ is the plasma velocity in the rotating frame.  In this reference frame the MRxMHD minimum energy states satisfy
\begin{align}
  \rho \left(\mathbf{u}' \cdot \nabla\right) \mathbf{u}' &= -\nabla p + \mathbf{J} \times \mathbf{B} + \rho \Omega_i^2 R \uR - 2 \rho \Omega_i \uZ \times \mathbf{u}', \label{eq:FiniteVolumeRotatingForceBalance}
\intertext{where the last two terms on the right-hand side are respectively the centrifugal and Coriolis forces.  This should be compared to the evolution equation for ideal MHD with flow in the same rotating reference frame, which is}
  \rho \frac{\partial}{\partial t} \mathbf{u}' + \rho (\mathbf{u}' \cdot \nabla) \mathbf{u}' &= - \nabla p + \mathbf{J} \times \mathbf{B} + \rho \Omega_i^2 R \uR - 2\rho \Omega_i \uZ \times \mathbf{u}'. \label{eq:IdealMHDEvolutionRotatingFrame}
\end{align}
A comparison of Eqs.~\eqref{eq:FiniteVolumeRotatingForceBalance} and \eqref{eq:IdealMHDEvolutionRotatingFrame} demonstrates that the minimum energy states of MRxMHD with flow satisfy $\frac{\partial}{\partial t} \mathbf{u}' = 0$.  As each plasma region is time-independent in a rotating reference frame, any 3D structures in plasma region $\mathcal{R}_i$ will rotate about the $Z$ axis with angular frequency $\Omega_i$ as seen in the laboratory frame.  Such rotating 3D structures are not just allowed by this model, but are actually realized for appropriate plasma constraints. \citet{Khalzov:2012} have recently studied the single-volume small-flow limit of this model in a cylinder, and in this limit they demonstrated the existence of minimum-energy states which are time-dependent in the laboratory frame.

We can now see from the force-balance equation, Eq.~\eqref{eq:IdealMHDEvolutionRotatingFrame}, that a contradiction would arise if we were to assume that toroidal angular momentum was conserved for non-axisymmetric plasma boundaries.  In such a case, $\Omega_i$ would not necessarily be zero, and therefore the 3D structure of the volume would rotate in time and intersect the fixed plasma boundary, which it cannot as the boundary is assumed to be impermeable.  The resolution is that the toroidal angular momentum is not conserved in a plasma with a non-axisymmetric outer boundary as, even in the absence of viscosity, forces exerted by the boundary normal to the wall can exert a non-zero torque on the plasma.

Finally, another important effect of flow on MRxMHD is that pressure is no longer constant in each plasma region.  In the absence of flow, the pressure profile of MRxMHD has a piecewise-constant structure \citep{Hole:2006}, but we see from Eq.~\eqref{eq:FiniteVolumeBernoulli} that this will not generally be the case when the effects of flow are included because the plasma flow velocity will vary throughout each plasma region according to Eq.~\eqref{eq:FiniteVolumeFlow}.

\section{The continuously nested flux-surface limit}
\label{sec:ContinuousLimit}

In this section we take the continuously nested flux surface limit ($N \rightarrow \infty$) of MRxMHD with flow and prove that it reduces to ideal MHD with flow. 

Taking the limit of infinitesimally small plasma regions of the energy functional Eq.~\eqref{eq:FiniteVolumeEnergyFunctional} gives
\begin{align}
  \begin{split}
  W &= \int \left(\frac{1}{2}\rho \mathbf{u}^2 + \frac{1}{2}\mathbf{B}^2 + \frac{1}{\gamma-1}\sigma(s) \rho^\gamma \right)d^3\tau\\
  &\relphantom{=} - \int \nu(s)\left(dM - dM^0\right) - \int\frac{1}{2}\mu(s)\left(dK - dK^0\right) - \int \lambda(s) \left(dC - dC^0\right) - \int \Omega(s)\left(dL - dL^0\right),
  \end{split}
  \label{eq:ContinuousEnergyFunctional1}
\end{align}
where $s$ is an arbitrary flux-surface label;  $dM$, $dK$, $dC$ and $dL$ are respectively infinitesimal amounts of plasma mass, magnetic helicity, flow helicity and toroidal angular momentum between infinitesimally separated flux surfaces; and $dM^0$, $dK^0$, $dC^0$ and $dL^0$ are the corresponding constraints.  The expressions for the infinitesimals $dM$, $dC$ and $dL$ follow immediately from Eqs.~\eqref{eq:NoFlowPlasmaMass}, \eqref{eq:FiniteVolumeFlowHelicity}, and \eqref{eq:FiniteVolumeAngularMomentum}, however the infinitesimal for the magnetic helicity $dK$ deserves additional attention.  From Eq.~\eqref{eq:NoFlowMagneticHelicity} we obtain that the infinitesimal of magnetic helicity $dK$ is
\begin{align}
  dK &= \mathbf{A}\cdot\mathbf{B}\, d^3\tau - \psi_t(s) d\psi_p + \psi_p(s) d\psi_t, \label{eq:InfinitesimalHelicity}
\end{align}
where $\psi_t(s)$ and $\psi_p(s)$ are respectively the toroidal and poloidal fluxes enclosed by flux surface $s$.  The enclosed magnetic fluxes are defined by
\begin{align}
  \psi_t(s) &= \oint_{\mathcal{C}_p(s)} \mathbf{A} \cdot d\mathbf{l}, \\
  \psi_p(s) &= -\oint_{\mathcal{C}_t(s)} \mathbf{A} \cdot d\mathbf{l},
\end{align}
with $\mathcal{C}_t(s)$ and $\mathcal{C}_p(s)$ being toroidal and poloidal circuits along the flux surface $s$.

\subsection{The magnetic flux constraints}
\label{sec:FluxConstraints}

In addition to the constraints listed in the energy functional Eq.~\eqref{eq:ContinuousEnergyFunctional1}, we must also enforce the magnetic flux constraints.  In the finite-volume case this is convenient to enforce using a relationship between the vector potential variation $\delta\mathbf{A}$ and the variation of the plasma interfaces $\delta\mathbf{x}$ (see Appendix~\ref{sec:MRxMHDDerivation}).  In the limit of continuously nested flux surfaces it is instead easier to enforce the flux constraints with Lagrange multipliers and adding the relevant terms to the energy functional Eq.~\eqref{eq:ContinuousEnergyFunctional1}.

In addition to the usual toroidal and poloidal flux constraints, in the limit of continuously nested flux surfaces the radial magnetic fluxes must be everywhere zero, consistent with the surfaces labeled by $s$ being magnetic flux surfaces.  The magnetic flux constraints have the form
\begin{align}
  \left.W\right|_{\text{flux constraints}} &= - 2\pi\int Q_\zeta(s) (d\psi_t - d\psi_t^0) - {2\pi}\int Q_\theta(s) (d\psi_p - d\psi_p^0) - \int Q_s(s,\theta,\zeta) \left(\mathbf{B} \cdot \nabla s\right) \, d^3\tau,
\end{align}
where $d\psi_t$ and $d\psi_p$ are infinitesimal elements of toroidal and poloidal magnetic flux respectively, and $Q_\zeta$ and $Q_\theta$ are the corresponding Lagrange multipliers.  The $Q_s$ Lagrange multiplier enforces the constraint $\mathbf{B}\cdot \nabla s = 0$.

The toroidal magnetic flux constraint can be expressed as a volume integral:
\begin{align}
  \int 2\pi Q_\zeta(s)\, d\psi_t &= 2\pi \int Q_\zeta(s) \mathbf{B}\cdot d^2\boldsymbol{\sigma}_\zeta = \left(\int d\zeta \right) \int Q_\zeta(s) \mathbf{B}\cdot\left(\mathbf{e}_s \times \mathbf{e}_\theta \right)\,ds\, d\theta = \int \mathbf{B} \cdot \left(Q_\zeta(s) \nabla \zeta\right) \, J\, ds\, d\theta\, d\zeta \\
  &= \int \mathbf{B}\cdot \left(Q_\zeta(s) \nabla \zeta\right) \,d^3\tau,
\end{align}
where $\mathbf{e}_s$, $\mathbf{e}_\theta$, $\mathbf{e}_\zeta$ are the covariant basis vectors \citep{FluxCoordinates} and $J$ is the Jacobian of the $(s, \theta, \zeta)$ coordinate system, with $\theta$ an arbitrary poloidal angle coordinate and $\zeta$ an arbitrary toroidal angle coordinate.  

Similarly the poloidal magnetic flux constraint can be expressed as a volume integral:
\begin{align}
  \int 2\pi Q_\theta(s)\, d\psi_p &= \int \mathbf{B} \cdot \left(Q_\theta(s) \nabla \theta\right) \, d^3 \tau.
\end{align}
Defining $\mathbf{Q} = Q_s \nabla s + Q_\theta \nabla \theta + Q_\zeta \nabla \zeta$ as the vector of Lagrange multipliers enforcing the magnetic flux constraints, the flux constraints can be written in the compact form
\begin{align}
  \left. W\right|_\text{flux constraints} &= - \int \left(\mathbf{Q} \cdot \mathbf{B}\right)\, d^3 \tau + 2\pi \int \left[Q_\theta(s) \frac{d\psi_p^0(s)}{ds} + Q_\zeta(s) \frac{d\psi_t^0(s)}{ds} \right]\,ds. \label{eq:ContinuousFluxConstraints}
\end{align}

\subsection{The magnetic helicity constraint}
The magnetic helicity constraints are trivially satisfied in the limit of infinitesimally separated magnetic flux surfaces.  We demonstrate this by expressing the gauge terms in Eq.~\eqref{eq:InfinitesimalHelicity} directly in terms of the $(s, \theta, \zeta)$ coordinate system:
\begin{align}
  -\psi_t\, d\psi_p &= -\left(\oint \mathbf{A}\cdot \mathbf{e}_\theta\,  d\theta\right)\left[\oint \mathbf{B} \cdot \left(\mathbf{e}_\zeta \times \mathbf{e}_s \right)\,d\zeta \right] \, ds = -\left(\oint A_\theta B^\theta \, J\, d\theta\, d\zeta\right)ds,\\
  \psi_p\, d\psi_t &= \left(-\oint \mathbf{A}\cdot \mathbf{e}_\zeta\, d\zeta\right)\left[\oint \mathbf{B} \cdot \left(\mathbf{e}_s \times \mathbf{e}_\theta\right)\, d\theta \right]\, ds  = -\left(\oint A_\zeta B^\zeta\, J\, d\theta\, d\zeta \right)ds,\\
  \implies dK &= \left[\oint \left(\mathbf{A}\cdot\mathbf{B} - A_\theta B^\theta - A_\zeta B^\zeta\right)\, J\, d\theta\, d\zeta \right]ds = 0,
\end{align}
where $\mathbf{A}\cdot\mathbf{B} = A_\theta B^\theta + A_\zeta B^\zeta + A_s B^s$ and $\mathbf{B}\cdot\nabla s = 0 \implies B^s = 0$.  In the limit that the interfaces $\mathcal{I}_i$ become continuously nested, the differential amount of magnetic helicity on each surface becomes zero, and the magnetic helicity constraint (which therefore must also be zero) is trivially satisfied.  Magnetic helicity behaves differently to other plasma quantities like plasma mass as the interfaces $\mathcal{I}_i$ become more closely separated because magnetic helicity is a topological quantity \citep{Berger:1999} and not an extrinsic quantity like plasma mass or an intrinsic quantity like plasma density.

\subsection{Variation of the energy functional of MRxMHD with flow}
\label{sec:VariationOfMRxMHDFlowEnergyFunctional}

Including the magnetic flux constraints given by Eq.~\eqref{eq:ContinuousFluxConstraints} and removing the trivially satisfied magnetic helicity constraint, the energy functional of Eq.~\eqref{eq:ContinuousEnergyFunctional1} becomes
\begin{align}
  \begin{split}
  W =& \int  \left[\frac{1}{2}\rho \mathbf{u}^2 + \frac{1}{2} \mathbf{B}^2 + \frac{1}{\gamma-1}\sigma(s)\rho^\gamma - \mathbf{Q}\cdot\mathbf{B} - \nu(s)\rho  - \lambda(s) \mathbf{B}\cdot\mathbf{u} - \rho \Omega(s) R \mathbf{u}\cdot\uphi \right] d^3\tau \\
  & + \int \left[2\pi Q_\theta(s) \frac{d\psi_p^0(s)}{ds} + 2\pi Q_\zeta(s) \frac{d\psi_t^0(s)}{ds} + \nu(s) \frac{dM^0(s)}{ds} + \lambda(s) \frac{dC^0(s)}{ds} + \Omega(s) \frac{dL^0(s)}{ds}\right]\, ds.
  \end{split} \label{eq:ReducedEnergyFunctional}
\end{align}
Variations of $W$ with respect to the Lagrange multipliers enforce the corresponding constraints.  The interesting variations are those with respect to $\rho$, $\mathbf{u}$, $\mathbf{B}$, and the position of the flux surfaces $\mathbf{x}$.

Setting the variations of $W$ with respect to $\rho$, $\mathbf{u}$ and $\mathbf{B}$ to zero yield respectively
\begin{align}
  \nu(s) &= \frac{1}{2} \mathbf{u}^2 + \frac{\gamma}{\gamma - 1}\sigma(s) \rho^{\gamma - 1} - \Omega(s)R \mathbf{u}\cdot\uphi, \label{eq:Bernoulli} \\
  \rho \mathbf{u} &= \lambda(s) \mathbf{B} + \rho \Omega(s) R \uphi, \label{eq:FlowFieldRelationship} \\
  \mathbf{Q} &= \mathbf{B} - \lambda(s) \mathbf{u}. \label{eq:QFieldRelation}
\end{align}
The first of these is Bernoulli's equation for ideal MHD (compare to Eq.~(31) of \citet{McClements:2001} with $V = 0$), and the second equation also appears in ideal MHD and is equivalent to Eq.~(29) of \citet{Hameiri:1998}.

The remaining variation of the energy functional $W$ with respect to the position of the flux surfaces $\mathbf{x}$ is
\begin{align}
  \left. \delta W \right|_{\delta \mathbf{x}} &= \int \left\{ \delta \mathbf{x} \cdot \left[-\frac{1}{\gamma-1}\rho^\gamma \nabla \sigma(s) + \rho \nabla \nu(s) + \mathbf{B}\cdot\mathbf{u} \nabla \lambda(s) + \rho R \mathbf{u}\cdot\uphi \nabla \Omega(s) \right]  - \left.\delta\mathbf{Q}\right|_{\delta \mathbf{x}} \cdot \mathbf{B} \right\}\, d^3\tau, \label{eq:DeltaWWithRespectToDeltaX}
\end{align}
where we have used $\delta s(\mathbf{x}) = - \delta \mathbf{x} \cdot \nabla s(\mathbf{x})$ (see Eq.~(17) of \citet{Dennis:2013a}), and $\left. \delta \mathbf{Q}\right|_{\delta\mathbf{x}}$ is the variation of $\mathbf{Q}(s,\theta,\zeta)$ due to the variation of the position of the flux surfaces.  

The  $\int \left. \delta\mathbf{Q}\right|_{\delta\mathbf{x}} \cdot \mathbf{B}\, d^3\tau$ term in Eq.~\eqref{eq:DeltaWWithRespectToDeltaX} can be simplified using the relation
\begin{align}
  \int \left. \delta\mathbf{Q}\right|_{\delta\mathbf{x}} \cdot \mathbf{B}\, d^3\tau &= \int \delta \mathbf{x} \cdot \left[ \left(\nabla \times \mathbf{Q}\right) \times \mathbf{B}\right] d^3\tau, \label{eq:deltaQSimplification}
\end{align}
which follows from expanding $\mathbf{Q} = Q_i(s,\theta,\zeta)\nabla u^i$ and using $\delta u^i(\mathbf{x}) = -\delta \mathbf{x} \cdot \nabla u^i$, where the $u^i$ are the magnetic coordinates $(s,\theta,\zeta)$ and we have used the Einstein summation convention over the repeated index $i$.

The variation of the energy functional with respect to $\delta \mathbf{x}$ can now be simplified using Eqs.~\eqref{eq:Bernoulli}, \eqref{eq:QFieldRelation} and \eqref{eq:deltaQSimplification} to
\begin{align}
  \left.\delta W \right|_{\delta \mathbf{x}}&= \int  \delta \mathbf{x} \cdot \left[-\mathbf{J}\times \mathbf{B} + \rho \left(\mathbf{u}\cdot\nabla\right)\mathbf{u} + \nabla p + \rho \Omega R \uphi \times \left(\nabla \times \mathbf{u}\right) - \rho \Omega \nabla \left(R \mathbf{u}\cdot \uphi\right) \right] \, d^3\tau. \label{eq:DeltaWWithRespectToDeltaX8}
\end{align}
Setting the variation $\left.\delta W \right|_{\delta \mathbf{x}}$ to zero gives
\begin{align}
  \rho \left(\mathbf{u} \cdot \nabla \right)\mathbf{u} &= -\nabla p + \mathbf{J} \times \mathbf{B} - \rho \Omega(s) R \uphi \times \left(\nabla \times \mathbf{u}\right) + \rho \Omega(s) \nabla (R \mathbf{u} \cdot \uphi). \label{eq:ContinuousSurfacesForceBalance1}
\end{align}

Comparing Eq.~\eqref{eq:ContinuousSurfacesForceBalance1} to the evolution equation for ideal MHD with flow, Eq.~\eqref{eq:IdealMHDEvolution}, demonstrates that the minimum energy state described by Eq.~\eqref{eq:ContinuousSurfacesForceBalance1} will not be time-independent unless the last two terms are zero.  As these terms depend on $\Omega(s)$, simplifying them will depend on the form of the angular momentum constraints assumed in the model.  As discussed in Section~\ref{sec:MRxMHDWithFlow}, the choice of angular momentum constraints applied to the model depend on the assumptions made and the geometry of the plasma boundary.  We consider three cases: (i) the plasma is assumed to remain axisymmetric during the relaxation process; (ii) only the outer boundary of the plasma is assumed to be axisymmetric, and the interior of the plasma may have 3D structure; and (iii) the outer boundary of the plasma is not axisymmetric.

\subsubsection*{Case 1: The plasma remains axisymmetric during plasma relaxation}

If the plasma remains axisymmetric during the relaxation process, then the toroidal angular momentum will be conserved on each flux surface and the Lagrange multiplier $\Omega(s)$ may vary across the plasma.  In this situation the last two terms of Eq.~\eqref{eq:ContinuousSurfacesForceBalance1} are zero
\begin{align}
  \rho \Omega(s) R \uphi \times \left(\nabla \times \mathbf{u}\right) - \rho \Omega(s) \nabla \left(R \mathbf{u}\cdot\uphi\right) &= -\rho \Omega(s) \left[ \hat{\mathbf{R}} \frac{\partial u_R}{\partial \phi} + \hat{\mathbf{Z}}\frac{\partial u_Z}{\partial \phi} + \uphi \frac{\partial u_\phi}{\partial \phi} \right] = 0, \label{eq:DeltaWAdditionalTerms}
\end{align}
and hence Eq.~\eqref{eq:ContinuousSurfacesForceBalance1} reduces to the force-balance equation for ideal MHD with flow
\begin{align}
  \rho \left(\mathbf{u}\cdot\nabla \right)\mathbf{u} &= -\nabla p + \mathbf{J} \times \mathbf{B}. \label{eq:IdealMHDForceBalance}
\end{align}
We have now proven that in axisymmetry the MRxMHD model defined by Eqs.~\eqref{eq:FiniteVolumePlasmaEnergy}--\eqref{eq:FiniteVolumeAngularMomentum} reduces to ideal MHD with flow as the number of plasma volumes $N$ becomes large.  In particular as the MRxMHD model in the axisymmetric limit (and with $N\rightarrow\infty$) reduces to the ideal MHD force-balance equation with the usual auxiliary equations Eqs.~\eqref{eq:Bernoulli}--\eqref{eq:FlowFieldRelationship}, the MRxMHD equilibria in the $N\rightarrow\infty$ limit will be described by the Grad-Shafranov equation with flow (see, for example, \citet[\S 18.2]{Goedbloed:2010}).

\subsubsection*{Case 2: Only the plasma boundary is assumed to be axisymmetric during relaxation}

If only the plasma boundary is assumed to be axisymmetric during the relaxation process then only the total toroidal angular momentum will be conserved and $\Omega(s)$ in Eq.~\eqref{eq:ContinuousSurfacesForceBalance1} should be replaced with the unknown scalar Lagrange multiplier $\Omega$ (see Section~\ref{sec:MRxMHDFlowEffects}).  In this case we find similarly to the results of Section~\ref{sec:MRxMHDFlowEffects} that the plasma is time-independent (and hence in force-balance) in a reference frame rotating with angular frequency $\Omega$ about the $Z$ axis.  Transforming Eq.~\eqref{eq:ContinuousSurfacesForceBalance1} into a rotating reference frame by making the replacement $\mathbf{u} \rightarrow \mathbf{u}' + \Omega R \uphi$ yields
\begin{align}
  \rho (\mathbf{u}' \cdot \nabla) \mathbf{u}' &= - \nabla p + \mathbf{J} \times \mathbf{B} + \rho \Omega^2 R \uR - 2 \rho \Omega \uZ \times \mathbf{u}',
\end{align}
which is the ideal MHD force-balance condition in the rotating reference frame (compare to Eq.~\eqref{eq:IdealMHDEvolutionRotatingFrame}).

By partially relaxing the usual assumption of axisymmetry of the plasma we have obtained an equilibrium model for 3D plasmas which rotate in the laboratory reference frame.  There is an additional restriction in comparison to axisymmetric ideal MHD with flow in that the usual flux function $\Omega(s)$ is now restricted to be the scalar Lagrange multiplier $\Omega$ which is constant across the entire plasma.

\subsubsection*{Case 3: The plasma boundary is not axisymmetric}

If the outer boundary of the plasma is not assumed to be axisymmetric, then angular momentum will not be conserved and we should make the replacement $\Omega(s)\rightarrow 0$ in Eq.~\eqref{eq:ContinuousSurfacesForceBalance1}.  In this case we again obtain the ideal MHD force-balance equation, Eq.~\eqref{eq:IdealMHDForceBalance}, but without any axisymmetry assumptions.

Fully relaxing the usual axisymmetry assumption yields an equilibrium model for 3D plasmas, but as the boundary is fixed and not axisymmetric, the flux function $\Omega(s)$ is now zero.  As a consequence of Eq.~\eqref{eq:FlowFieldRelationship}, in this limit the plasma flow is aligned with the magnetic field and given by $\rho \mathbf{u} = \lambda(s) \mathbf{B}$.

\subsubsection*{Summary}

We have now proven that as the number of plasma regions $N$ in MRxMHD with flow becomes large that the model reduces to ideal MHD with flow, either in a rotating reference frame or in the laboratory reference frame depending on the symmetry assumptions made in the model.

In the next section we apply MRxMHD with flow to a simple RFP-like plasma with flow.

\section{Example application to RFP-like plasma with flow}
\label{sec:Example}

In this section we apply our MRxMHD with flow model to an RFP-like plasma with a small amount of flow.  We take the small-flow limit for convenience of analytic calculations, the large flow limit would require numerically solving the nonlinear system of equations \eqref{eq:FiniteVolumeField}--\eqref{eq:InterfaceForceBalance}.

Our example application of MRxMHD is motivated by the experimental results of \citet{Kuritsyn:2009}.  In their work \citeauthor{Kuritsyn:2009} measured the parallel flow in the Madison Symmetric Torus (MST) reversed field pinch during a reconnection event and found that the normalized parallel momentum density $\rho \mathbf{u}\cdot \mathbf{B} / \mathbf{B}^2$ was roughly constant in the plasma core, but changed signs near the edge of the plasma.  This experiment has been studied by \citet{Khalzov:2012} who demonstrated that the experiment could be modeled as a plasma in a relaxed single-fluid MHD state.  This model is identical to the MRxMHD model with flow presented in this work in the limit of a single plasma volume and weak plasma flow.  The example model we present here is a minor extension of the work of \citet{Khalzov:2012} to include two plasma volumes to better describe the change in parallel momentum density near the edge of the plasma.

Motivated by the model of \citet{Khalzov:2012}, we approximate the MST as a periodic cylinder and consider the limit of weak plasma flow ($\rho \mathbf{u}^2 \ll \mathbf{B}^2$) and purely field-aligned flow ($\Omega = 0$).  To first order in $\lambda$, the plasma in each region satisfies
\begin{align}
  \nabla \times \mathbf{B} &= \mu_i \mathbf{B} + \lambda_i \nabla \times \mathbf{u}, \\
  \rho \mathbf{u} &= \lambda_i \mathbf{B}, \label{eq:RFPExampleFieldAlignedFlow}\\
  \nu_i &= \frac{\gamma}{\gamma - 1}\sigma_i \rho^{\gamma - 1},
\end{align}
where $\Omega_i=0$.  In this limit the plasma has a uniform pressure and density in each plasma region.

\begin{figure}
  \includegraphics[width=18cm]{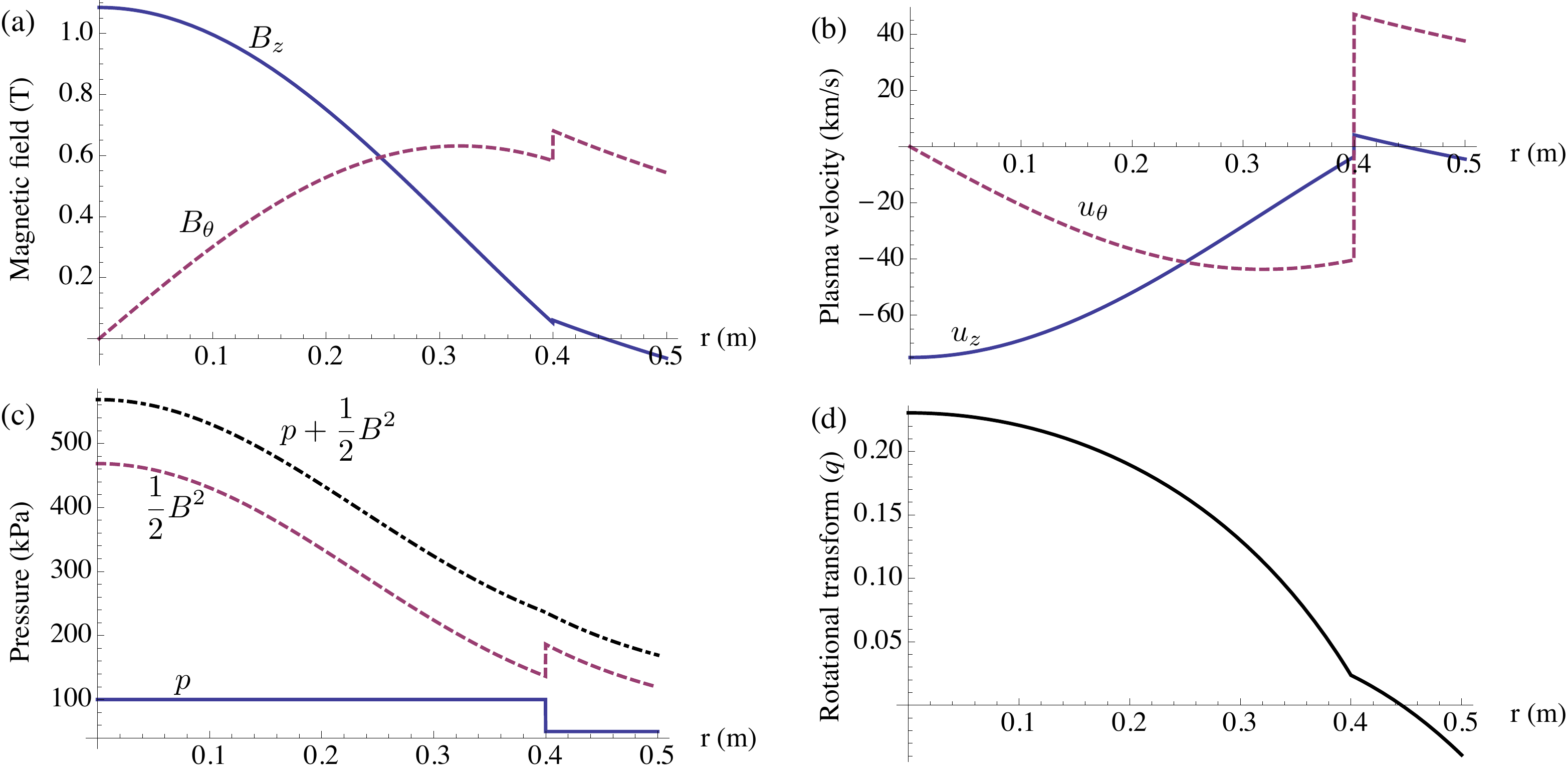}
  \caption{\label{fig:RFPExample}Example MRxMHD solution for an RFP in cylindrical geometry with two plasma volumes.  Panels (a) and (b) respectively show the magnetic field and plasma velocity components versus radial position.  Panel (c) shows the plasma pressure $p$, magnetic pressure $\frac{1}{2} B^2$ and total pressure $p + \frac{1}{2} B^2$ across the plasma.  Panel (d) shows the plasma rotational transform profile.}
\end{figure}

Figure~\ref{fig:RFPExample} shows an example RFP-like plasma with two plasma volumes.  The equilibrium is described by $\mu_1 = 5.79\,\text{m}^{-1}$, $\mu_2 = 2.04\,\text{m}^{-1}$, $r_1 = 0.4\,\text{m}$, $r_2 = 0.5\,\text{m}$ $p_1 = 100\,\text{kPa}$, $p_2 = 50\,\text{kPa}$, $\lambda_1/\sqrt{\rho_1} = -10^{-2}$, $\lambda_2/\sqrt{\rho_2} = +10^{-2}$, $\rho_1 = \rho_2 = 1.7 \times 10^{-8}\,\text{kg}/\text{m}^3$.  These values have been chosen to match the experimental plasma parameters \citep{Kuritsyn:2009} ($F=-0.2$, $\Theta = 1.7$, enclosed toroidal flux $\Phi_z \approx 40\,\text{mWb}$, plasma number density $n \approx 10^{19}\,\text{m}^{-3}$, ${n \mathbf{u}\cdot\mathbf{B}}/{\mathbf{B}^2} \approx \mp 5 \times 10^{20}\, \text{m}^{-3}\, \text{km}\, \text{s}^{-1}\, \text{T}^{-1}$).  Despite the discontinuous pressure profile the ideal MHD transport barrier at $r_1 = 0.4\,\text{m}$ is in force balance, as demonstrated in Figure~\ref{fig:RFPExample}(c) which shows that $p+\frac{1}{2}B^2$ is continuous across the interface, and hence the interface condition Eq.~\eqref{eq:InterfaceForceBalance} is satisfied.  The strong discontinuity in the poloidal plasma flow illustrated in Fig.~\ref{fig:RFPExample}(b) is due to the large jump in $\lambda_i$ at the reversal surface.  This feature derives from the large change in the parallel momentum of the plasma near the reversal surface in the experiment of \citet{Kuritsyn:2009}, and is illustrated in their Fig.~4(b), which is effectively a plot of our $\lambda$.

The example presented in this section demonstrates the existence of multi-volume, RFP-like solutions to the MRxMHD model with flow.  By increasing the number of interfaces, the plasma can be approximated arbitrarily close to any ideal MHD equilibrium, as proven in Section~\ref{sec:ContinuousLimit}.

\section{Conclusion}
\label{sec:Conclusion}

We have formulated an energy principle for equilibria that comprise multiple Taylor-relaxed plasma regions including the effects of plasma flow.  This model is an extension of earlier work that considered the zero-flow limit \citep{Hole:2007,Hole:2009a,Dennis:2013a} and the single relaxed-region limit \citep{Woltjer:1958,Finn:1983}.  We have demonstrated our model reduces to ideal MHD with flow in the limit of an infinite number of plasma regions.  In this limit, the magnetic geometry is characterized by continuously nested flux surfaces.  However the appeal of MRxMHD with flow is that the model is well-defined for flowing 3D plasmas as only a finite number of flux surfaces are assumed to exist.  The rest of the plasma may be characterized by smoothly nested flux surfaces, islands, chaotic fields or some combination of these.  The numerical solution to MRxMHD with flow in the nonlinear 3D case will be the subject of future work as an extension to the Stepped Pressure Equilibrium Code \citep{Hudson:2012}. A unique feature of the model presented here is that it allows an energy-minimization approach to be used to the description of plasmas with rotating 3D structure such as the long-lived mode on MAST \citep{Chapman:2010,Hua:2010} or the `snake' on various devices \citep{Weller:1987,Pecquet:1997,Menard:2005,Delgado-Aparicio:2013}.  

The authors gratefully acknowledge support of the U.S. Department of Energy and the Australian Research Council, through Grants No.\ DP0452728, No.\ FT0991899, and No.\ DP110102881.

\appendix

\section{Derivation of the MRxMHD equations}
\label{sec:MRxMHDDerivation}

In this appendix we derive the MRxMHD equations for the plasma, Eqs.~\eqref{eq:FiniteVolumeField}--\eqref{eq:FiniteVolumeBernoulli}, and the interface condition, Eq.~\eqref{eq:InterfaceForceBalance}.  The plasma equations \eqref{eq:FiniteVolumeField}--\eqref{eq:FiniteVolumeBernoulli} have been obtained previously by \citet{Woltjer:1958} and \citet{Finn:1983} in the context of single relaxed-region models.  The derivation is essentially unchanged when considering the case of a finite number of nested relaxed-regions, which is the case considered here.  The new result presented here is the interface condition Eq.~\eqref{eq:InterfaceForceBalance}.  We also present a derivation of the plasma equations for completeness and as a necessary step in obtaining the interface condition.

Equilibria of the MRxMHD model are stationary points of the energy functional Eq.~\eqref{eq:FiniteVolumeEnergyFunctional},
\begin{align}
  W &= \sum_i E_i - \sum_i \nu_i(M_i - M_i^0) - \frac{1}{2}\sum_i \mu_i(K_i - K_i^0) - \sum_i \lambda_i (C_i - C_i^0) - \sum_i \Omega_i (L_{i} - L_{i}^0), \label{eq:FiniteVolumeEnergyFunctionalAppendix}
\end{align}
where $\nu_i$, $\mu_i$ and $\Omega_i$ are Lagrange multipliers and $E_i$, $M_i$, $K_i$, $C_i$ and $L_{i}$ are defined by Eqs.~\eqref{eq:FiniteVolumePlasmaEnergy}--\eqref{eq:FiniteVolumeAngularMomentum}.  

Instead of introducing Lagrange multipliers to enforce the toroidal and poloidal flux constraints as in Section~\ref{sec:FluxConstraints}, we use the approach of \citet{Spies:2001} who showed that the flux constraints are equivalent to the following relationship at the interfaces
\begin{align}
  \mathbf{n} \times \delta\mathbf{A} &= - \left(\mathbf{n}\cdot\delta\mathbf{x}\right) \mathbf{B}, \label{eq:DeltaAInterfaceConstraint}
\end{align}
where $\delta\mathbf{A}$ is the variation of the vector potential and $\delta\mathbf{x}$ is the perturbation to the interface positions.

Setting the variations of $W$ with respect to $\mathbf{u}$ and $\rho$ to zero yield respectively
\begin{align}
  \rho \mathbf{u} &= \lambda_i \mathbf{B} + \rho \Omega_i R \uphi, \\
  \nu_i &= \frac{1}{2} \mathbf{u}^2 + \frac{\gamma}{\gamma - 1} \sigma_i \rho^{\gamma - 1} - \Omega_i R \mathbf{u} \cdot \uphi,
\end{align}
which are the last two plasma bulk conditions, Eqs.~\eqref{eq:FiniteVolumeFlow} and \eqref{eq:FiniteVolumeBernoulli}.

The variation of $W$ with respect to $\mathbf{A}$ is
\begin{align}
  \left. \delta W \right|_{\delta\mathbf{A}} &= \sum_i\int_{\mathcal{R}_i} \delta\mathbf{A} \cdot \left(\nabla \times \mathbf{B} - \lambda_i \nabla \times \mathbf{u} - \mu_i \mathbf{B} \right) + \sum_i\oint_{\partial \mathcal{R}_i} \left(\mathbf{n} \times \delta\mathbf{A}\right) \cdot \left(\mathbf{B} - \lambda_i \mathbf{u} - \frac{1}{2}\mu_i \mathbf{A}\right)\, d^2\sigma,
\end{align}
where $\mathbf{n}$ is a unit normal perpendicular to the boundary of the plasma volume, $\partial\mathcal{R}_i = \mathcal{I}_{i-1} \cup \mathcal{I}_i$ is the boundary of the plasma volume $\mathcal{R}_i$, and $\mathcal{I}_i$ is the plasma interface separating plasma volumes $\mathcal{R}_{i-1}$ and $\mathcal{R}_i$ (see Figure~\ref{fig:NestedSurfaces}).  Using Eq.~\eqref{eq:DeltaAInterfaceConstraint}, the surface integral in $\left.\delta W\right|_{\delta\mathbf{A}}$ can be written in terms of the variation to the plasma interfaces $\delta\mathbf{x}$
\begin{align}
  \left. \delta W \right|_{\delta\mathbf{A}} &= \sum_i\int_{\mathcal{R}_i} \delta\mathbf{A} \cdot \left(\nabla \times \mathbf{B} - \lambda_i \nabla \times \mathbf{u} - \mu_i \mathbf{B} \right) - \sum_i\oint_{\partial \mathcal{R}_i} \left(\mathbf{n}\cdot \delta\mathbf{x}\right) \left(\mathbf{B}^2 - \lambda_i \mathbf{u}\cdot\mathbf{B} - \frac{1}{2}\mu_i \mathbf{A}\cdot\mathbf{B}\right) \, d^2\sigma. \label{eq:DeltaWiDeltaA}
\end{align}
Requiring $\left.\delta W\right|_{\delta\mathbf{A}}$ to be zero for all choices of $\delta\mathbf{A}$ yields
\begin{align}
  \nabla \times \mathbf{B} &= \mu_i \mathbf{B} + \lambda_i \nabla \times \mathbf{u},
\end{align}
which is the first plasma bulk condition, Eq.~\eqref{eq:FiniteVolumeField}.

The interface condition can now be obtained by considering the variation of $W$ with respect to the interface positions
\begin{align}
  \begin{split}
  \left. \delta W \right|_{\delta \mathbf{x}} =& \sum_i\oint_{\partial \mathcal{R}_i} \left(\mathbf{n} \cdot \delta\mathbf{x}\right) \left( \frac{1}{2} \rho \mathbf{u}^2 + \frac{1}{2} \mathbf{B}^2 + \frac{1}{\gamma - 1}\sigma_i \rho^\gamma - \nu_i \rho - \lambda_i \mathbf{B}\cdot\mathbf{u} - \rho \Omega_i R \mathbf{u}\cdot \uphi - \frac{1}{2}\mu_i \mathbf{A}\cdot\mathbf{B}\right) \\
  &- \sum_i\oint_{\partial \mathcal{R}_i} \left(\mathbf{n}\cdot\delta\mathbf{x}\right) \left(\mathbf{B}^2 - \lambda_i \mathbf{u}\cdot\mathbf{B} - \frac{1}{2}\mu_i \mathbf{A}\cdot\mathbf{B}\right),
  \end{split} \label{eq:DeltaWiDeltaXIntermediate}
\end{align}
where the remaining term of Eq.~\eqref{eq:DeltaWiDeltaA} has been included.  Equation~\eqref{eq:DeltaWiDeltaXIntermediate} simplifies to
\begin{align}
  \left. \delta W \right|_{\delta \mathbf{x}} &= - \sum_i \oint_{\mathcal{I}_i} \left(\mathbf{n}\cdot\delta\mathbf{x}\right) \left[\left[ p + \frac{1}{2} \mathbf{B}^2 \right]\right],
\end{align}
where $\left[\left[x_i\right]\right] = x_{i+1} - x_{i-1}$ is the jump in $x$ across the plasma interface $\mathcal{I}_i$.  Requiring this variation to be zero gives the interface condition Eq.~\eqref{eq:InterfaceForceBalance},
\begin{align}
  \left[\left[ p + \frac{1}{2} \mathbf{B}^2 \right]\right] &= 0
\end{align}

\bibliography{MRXMHD_Flow_Limit_Paper.bib}

\end{document}